\documentclass[twocolumn,showpacs,superscriptaddress,floatfix,aps,prl]{revtex4-1}
\usepackage{amsmath}
\usepackage[dvips]{graphicx}
\usepackage{color}
\definecolor{blue}{rgb}{0,0,1.0}
\definecolor{green}{rgb}{0,1.0,0}
\definecolor{red}{rgb}{1.0,0,0}

\DeclareTextSymbol{\degre}{T1}{6}
\DeclareTextSymbol{\degre}{OT1}{23}

\begin{document}

\author{M. Dubois}
\affiliation{Institut Langevin, ESPCI ParisTech CNRS UMR7587, 1 rue Jussieu, 75238 Paris cedex 05, France}

\author{E. Bossy}
\affiliation{Institut Langevin, ESPCI ParisTech CNRS UMR7587, 1 rue Jussieu, 75238 Paris cedex 05, France}

\author{S. Enoch}
\affiliation{Aix-Marseille Universit\'{e}, CNRS, Centrale Marseille, Institut Fresnel, UMR 7249, 13013 Marseille, France}

\author{S. Guenneau}
\affiliation{Aix-Marseille Universit\'{e}, CNRS, Centrale Marseille, Institut Fresnel, UMR 7249, 13013 Marseille, France}

\author{G. Lerosey}
\affiliation{Institut Langevin, ESPCI ParisTech CNRS UMR7587, 1 rue Jussieu, 75238 Paris cedex 05, France}

\author{P. Sebbah}
\email[Contact: ]{patrick.sebbah@espci.fr}
\affiliation{Institut Langevin, ESPCI ParisTech CNRS UMR7587, 1 rue Jussieu, 75238 Paris cedex 05, France}

\date{\today}

\title{Time Driven Subwavelength Focusing with Negative Refraction}

\begin{abstract}
Flat lens concept based on negative refraction proposed by Veselago in 1968 has been mostly investigated in monochromatic regime. It was recently recognized that time development of the super-lensing effect discovered in 2000 by Pendry is yet to be assessed and may spring surprises: Time-dependent illumination could improve the spatial resolution of the focusing. We investigate dynamics of flexural wave focusing by a 45\degre-tilted square lattice of circular holes drilled in a Duraluminium plate. Time-resolved experiments reveal that the focused image shrinks with time below diffraction limit, with a lateral resolution increasing from 0.8 $\lambda$ to 0.35 $\lambda$, whereas focusing under harmonic excitation remains diffraction limited. Modal analysis reveals the role in pulse reconstruction of radiating lens resonances, which repeatedly self-synchronize at the focal spot to shape a super-oscillating field.
\end{abstract}

\maketitle

A conventional lens that focuses a single color may not necessarily focus a short broadband pulse: Chromatic and spherical aberration together with radially varying group delay are examples of dispersive effects which distort the pulse both in space and time and make its reconstruction in the image plane challenging. At the interface of a negative refractive index medium, waves can bend to a negative angle. Negative refraction (NR) was first envisioned by Veselago \cite{Veselago68} to design a flat lens without optical axis and aperture restriction \cite{Parimi03}: A diverging point source comes into focus inside the lens before diverging and converging again to a point in free space at the second interface. Pendry predicted sub-wavelength focussing of the energy as a result of evanescent wave enhancement within the lens \cite{Pendry00}. Flat lens focusing has been demonstrated for electromagnetic \cite{SoukoulisPRL03,Parimi03,Eleftheriades04}, optic \cite{Berrier04,Schonbrun06,Matsumoto07,Lu07,Lippens08} and acoustic waves \cite{Sheng04,PagePRL11,PagePRB11,Burgholzer12,Dubois}. However, experimental investigation of flat lens focusing has been restricted to monochromatic or narrowband excitations, both in negative index metamaterials and photonic/phononic crystals \cite{Joannopoulos02}. That a time-varying signal could be negatively refracted has been a matter of debate \cite{Valanju02,SmithPendry02,Pacheco02,PendrySmith03}. Pulse reconstruction by negative refraction through a flat lens has been addressed only recently at least theoretically \cite{Gomez-Santos,Chan05,Pendry11,Greffet}. Time-dependent illumination is expected to improve spatial resolution of the focusing \cite{Chan05,Greffet,Heber}: Surface modes with increasing quality (Q) factors, which contribute to evanescent wave enhancement, should separate in time as high-Q surface modes would take longer to establish and to release their stored-energy \cite{Pendry11}. This prediction has not yet been tested experimentally.

Sub-wavelength resolution based on a different mechanism was recently demonstrated in metallic diffraction gratings: rapid changes of the field, known as super-oscillations, are observed in the far-field without contributions from evanescent fields, by precisely tailoring the interference of waves emitted by the grating \cite{ZheludevAPL07, ZheludevNano09,ZheludevJOpt13}. Berry and Popescu showed that these super-oscillations propagate over much greater distances than evanescent features \cite{Berry&Popescu}. Focusing is therefore not restricted to near-field excitation and imaging, making possible an optical microscope with super-resolution \cite{EleftheriadesSR13,ZheludevNatMat12}. However, the design of super-oscillatory function corresponding to realistic parameters remains challenging \cite{Toraldo,Aharonov,Berry&Popescu}.

In this letter, we investigate experimentally the dynamics of flat lens focusing for elastic waves propagating in a thin plate. It was shown recently that a properly designed array of see-through holes drilled in Duraluminium plate makes an efficient flat lens to refocus bending waves emanating from a point source \cite{Dubois}. Here we use this system to follow the time evolution of a short elastic pulse. We observe a  significant decrease with time of the lateral resolution of the focal spot, as predicted in \cite{Chan05,Greffet}. A resolution of 0.35 $\lambda$ is achieved with broadband excitation, while continuous wave (CW) excitation in this geometry never focuses below diffraction limit. Spectral analysis in the reciprocal space reveals how resonances progressively release energy stored in the lens. We show that lens modes excited by evanescent waves do not play any role in the refocusing. In contrast, modes that re-emit propagating waves contribute coherently to a time-evolving interference pattern at the image plane. As time progresses, amplitudes and phases self-adjust to spontaneously yield a super-oscillating field which oscillates faster than $\lambda/2$. 3D elastic finite difference time domain (FDTD) simulations confirms this mechanism and show quantum revival of the super-oscillatory focusing as predicted in \cite{Berry&Popescu}.

Bending waves are dispersive elastic waves propagating in plates with thickness much smaller than the wavelength. Broadband excitations and time domain measurements are the standard in acoustics. Time domain investigation is further facilitated as the human scale flat geometry is naturally well-adapted to point source excitation and direct optical imaging of the elastic field everywhere within the plate. A 45\degre-tilted square lattice (15 mm lattice constant) of see-through holes forms a 234 mm $\times$ 98 mm rectangular flat lens at the center of a 2mm-thick duraluminium rectangular plate. The lattice and holes have been dimensioned to maximize large angle negative refraction \cite{Dubois}. A Gaussian pulse centered at $f_0$ = 9 kHz with a bandwidth $\sigma$ = 2.1 kHz is synthesized and launched by a small acoustic piezo-actuator 18 mm away from the lens. At this frequency, elastic wavelength, $\lambda$ = 47.1 mm, is several times the plate thickness, which guaranties that only the fundamental cutoff-free anti-symmetric and symmetric modes, $A_0$ and $S_0$, are present in the plate. A laser velocimeter scans the acoustic velocity field, $v(\vec{r},t)=A(\vec{r},t)\cos(2\pi f_0t+\phi(\vec{r},t))$, of the out-of-plane component $A_0$. A more detailed description of the system and the experimental setup is given in \cite{Dubois}.

\begin{figure}[ht!]
\begin{center}
  \includegraphics[scale=0.55]{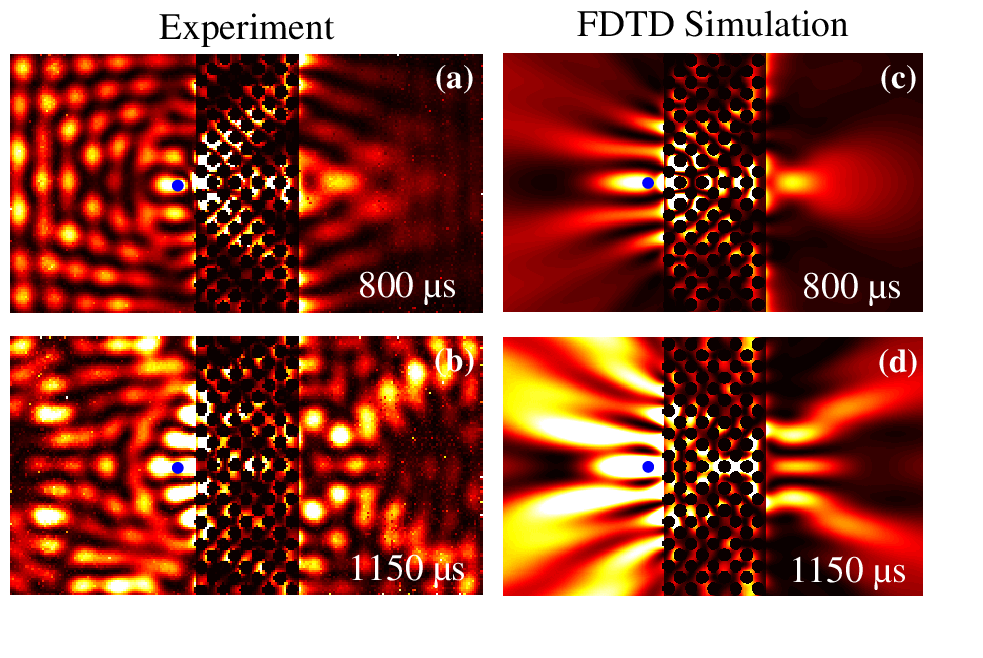}
  \caption{\label{Fig1}
      (Color online) Snapshots of the field intensity of vertical-displacement velocity measured at the surface of the plate at $t$= 800 $\mu$s (a,c) and $t$= 1150 $\mu$s (b,d). (a,b): Experiment; (c,d): Numerical simulations using 3D elastic FDTD with absorbing boundary conditions at the plate's edges. The color scale outside the crystal is normalized at each time step to the maximum intensity on the right side of the lens. Color scale has been reduced six times inside the crystal to avoid image saturation. Blue dot: Source position.
  }
\end{center}
\end{figure}

\begin{figure}[ht!]
\begin{center}
\includegraphics[scale=0.42]{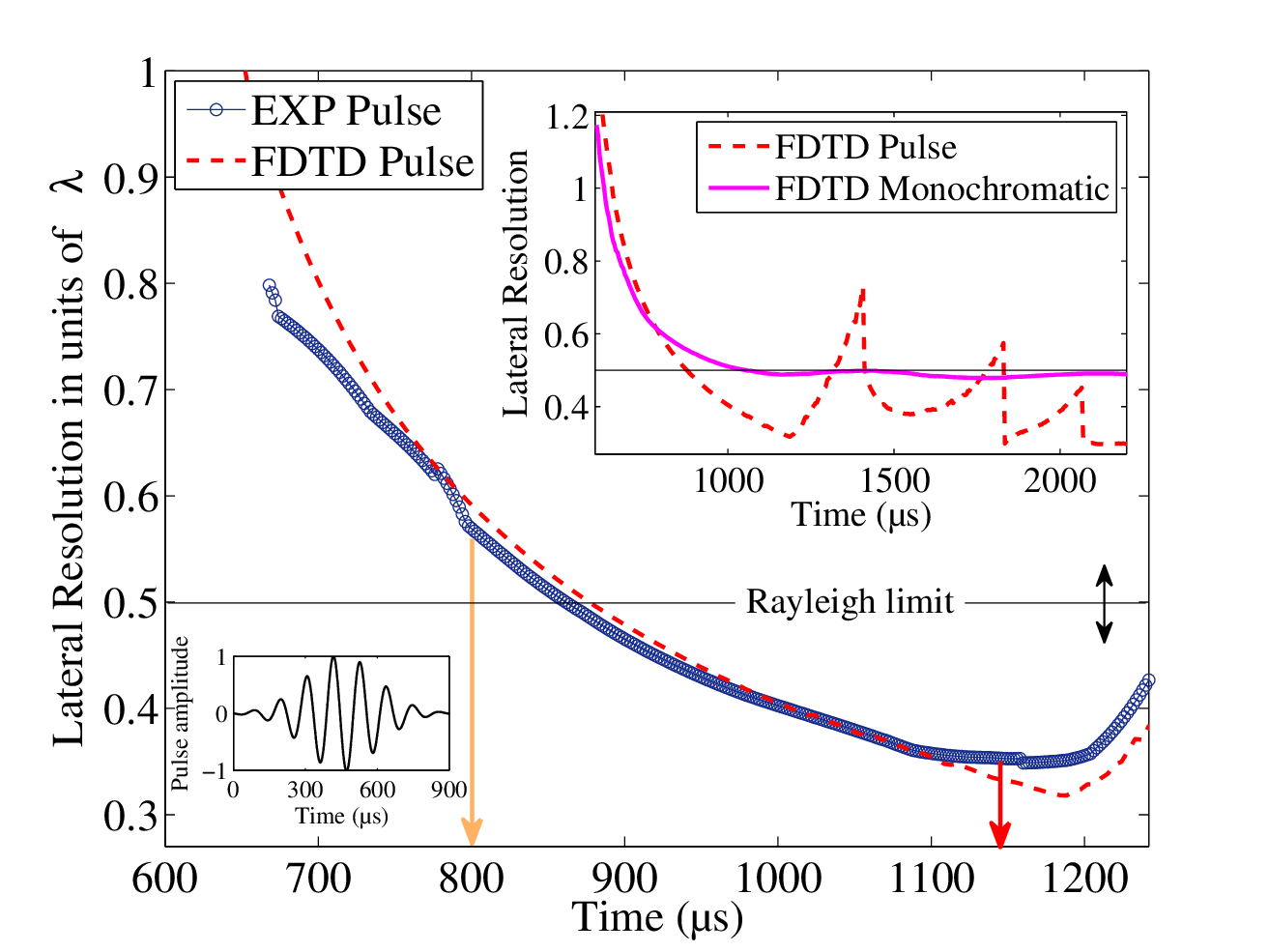}
\caption{\label{Fig2}
    (Color online) Time dependence of the lateral resolution in units of $\lambda$ ($\lambda$ = 47.1 mm at 9 kHz). Open circles: Experimental data; Dashed red line: Numerical simulations. Rayleigh diffraction limit is marked by the horizontal line at 0.5 $\lambda$. In actuality, this limit is ill-defined for broadband pulse excitation. The actual extension of the Rayleigh limit is indicated by the double vertical arrow and corresponds to $\pm$ 0.03 $\lambda$ for a pulse linewidth of $\sigma$ = 2.1 kHz, shown in lower inset. Vertical arrows correspond to the two time steps shown in Fig~\ref{Fig1}. Upper inset: larger time scale (numerical simulations). Solid magenta line: Monochromatic excitation. Dashed red line: Pulse excitation.
}
\end{center}
\end{figure}

The slowly varying field intensity, $|A(\vec{r},t)|^2$ is shown in Fig.~\ref{Fig1}a and b at $t$ = 800 $\mu$s and 1150 $\mu$s. In this figure, the field intensity has been renormalized at time $t$ by the averaged field intensity at focal point to compensate for the pulse extinction. The time evolution of the field intensity (without normalization) is presented at all times in a movie \cite{SM}. The pulse is seen to converge at the image plane to a spot which sharpens as time progresses. The shrinkage of the spot predicted in \cite{Pendry11,Greffet} is directly observed in our experiment. This dynamical feature is accompanied by the displacement of the focal point, another feature never discussed in the context of the NR flat lens and is reminiscent of the dispersion effects observed in regular lenses. To measure the lateral resolution of the refocusing, it is therefore necessary to track the position of the spot maximum versus time. The time evolution of the lateral resolution given by the full width at half maximum of the lateral profile at the focal plane, is shown in Fig.~\ref{Fig2}a. The resolution improves from 0.8 $\lambda$ to 0.35($\pm$0.03) $\lambda$, significantly below the diffraction limit with the minimum reached at time $t_{\rm focus}$ = 1176 $\mu$s. We find that this effect is maximum for $f_0$ = 9 kHz. However, similar resolution improvement is also observed for pulse excitation with carrier frequency ranging from 8 to 11 kHz. Noteworthily, as the resolution improves, side-lobes increase as predicted in \cite{Greffet}. The contrast is however preserved in the vicinity of the focus, offering the possibility of spatial filtering for imaging application. Note the field intensity distribution within the lens (Fig.~\ref{Fig1}b): a focal spot appears within the lens and Veselago ray picture is recovered. This has never been directly observed so clearly in an actual experiment, the nearest result we are aware of being \cite{Lippens08}.

We perform 3D elastic FDTD simulations \cite{Virieux,Bossy}. Highly absorbing boundary layers \cite{Graves} at the plate's edges totally remove the standing waves observed in the experiment (Fig.~\ref{Fig1}). Time driven improvement of the spatial resolution is confirmed with an almost perfect coincidence between experiment and simulations (Fig.~\ref{Fig2}): The effect is slightly more pronounced in the simulation, with a spot shrinkage from 1.2 $\lambda$ to 0.32 $\lambda$. At this point, it is interesting to compare refocusing of a pulse with refocusing of a step-like excitation where the amplitude of the oscillations at 9 kHz is kept constant after a gaussian transient rise of the field intensity. This is tested numerically since in the experiment spurious reflections at the edges of the plate would result in standing waves in the CW regime. During the transient regime, a focal spot forms in the image plane of the flat lens which reduces from 1.2 $\lambda$ down to 0.49 $\lambda$. The CW spot remains diffraction-limited as shown in dotted line in the upper inset of Fig.~\ref{Fig2}, in contrast the pulsed regime. Interestingly enough, the possibility to enhance the spatial resolution by considering time-domain excitation was certainly missed in earlier experimental and numerical studies, as they concentrated on the monochromatic regime. Time plays an unexpected role in pulse reconstruction that we now wish to investigate.

\begin{figure}[ht!]
		        \centering
			 \includegraphics[scale=0.65]{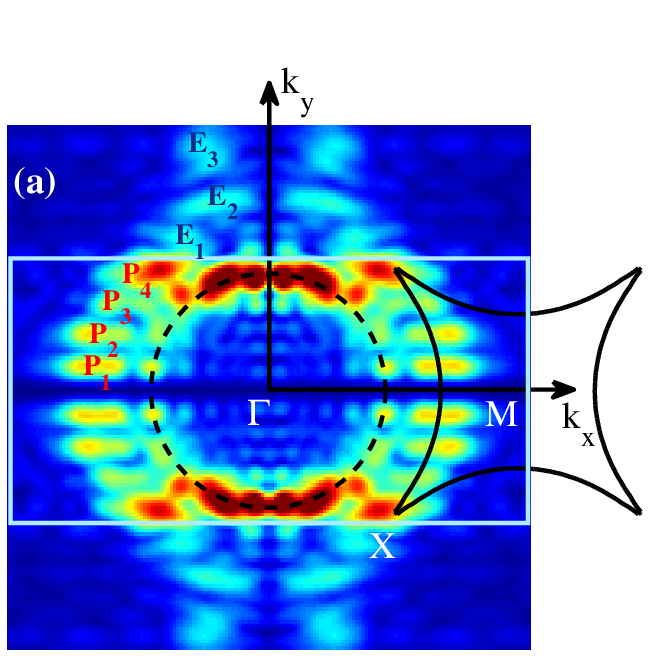}
			\centering
        		\includegraphics[scale=0.65]{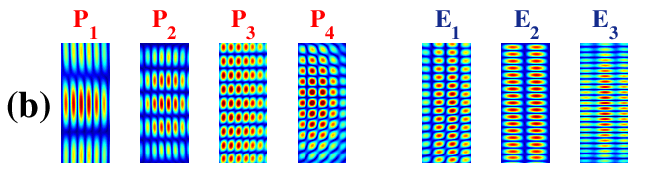}
	\caption{\label{Fig3}
    (Color online). (a) Snapshot of the reciprocal space at time $t_{\rm focus}$=1176 $\mu$s. $\Gamma$, X, M are the symmetry points of the square lattice and $k_x$ and $k_y$ the reciprocal array axes. Dashed line: 9 kHz-isofrequency isotropic contour in``free space'' (plate outside the platonic lens). Solid line: diamond-shaped 9 kHz-isofrequency contour within the platonic crystal. The white rectangle delimits the area of free-space propagative waves. The bright spots correspond to the modes of the flat lens selected spatially and spectrally by the point source. (b) Spatial amplitude distribution of the modes in the flat lens, obtained by inverse Fourier transform of the bright spots.
            }
\end{figure}

\begin{figure}[ht!]
\begin{center}
  \includegraphics[scale=0.45]{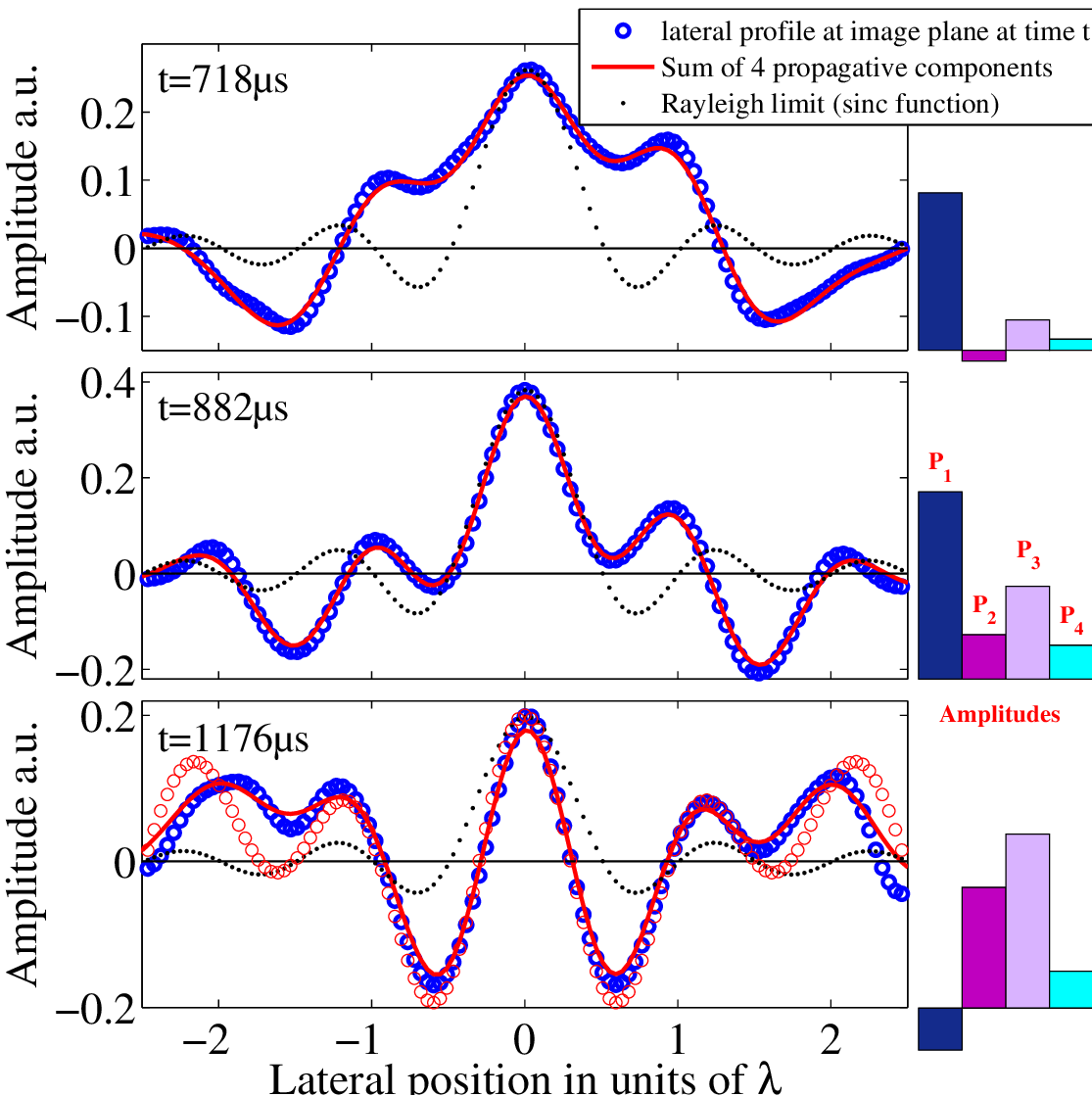}
  \caption{\label{Fig4}
    (Color online). Lateral profiles in the focal plane at three different time steps. Blue circles: Experimental data; Red solid line: Coherent superposition of the four propagative components radiated by modes $P_1$ to $P_4$ of Fig.~\ref{Fig3}. Black dots: Rayleigh diffraction limit (sinc function). Histograms show the contribution of propagative components at focal spot ($P_1$ to $P_4$ from left to right). Red circles: Super-oscillating function given by the sum of four weighted cosine functions (see text).
  }
\end{center}
\end{figure}

To address this issue, we carry out a modal analysis of the open-cavity formed by the flat lens and explore the modal interplay as time progresses. Both evanescent and propagating waves radiating from the source couple resonantly to discrete modes of the finite-sized platonic crystal. Each mode in turn releases its stored energy and a spot forms in the image plane as a coherent superposition of radiating fields. These short-lived modes are difficult to investigate because they overlap spectrally and spatially. We therefore perform a 2D spatial Fourier analysis of the field distribution at each time step. After applying a low-pass filter, the slowly-varying intensity in k-space is obtained. Because modes have different wavevectors, they can be distinguished in the k-space and followed dynamically. Figure \ref{Fig3}a shows the first Brillouin zone of the reciprocal space at $t_{\rm focus}$. The exchange of energy between the homogeneous plate and the platonic crystal, and the progressive excitation of the discrete modes of the lens are illustrated in a movie of the time-dependent evolution of the k-space energy distribution \cite{SM}. Superimposed in Fig.~\ref{Fig3}a are the isofrequency contours for the uniform plate (dashed circle) and for the infinite phononic crystal (solid-line diamond) at 9 kHz. The structural anisotropy of the crystal translates into non-isotropic an iso-contour around the $M$ point, in contrast to the circular contour around the $\Gamma$ point for the plain plate where propagation is isotropic. The complete calculation of the band diagram and the isofrequency contours for the infinite lattice can be found in \cite{Dubois}. Because the isofrequency contours shrink around the $M$ point with increasing frequency above 8 kHz, the group velocity points inward, bending diverging waves back to a focus \cite{Dubois}. Note that the spectral content of the pulse is broad and that energy distribution extends beyond these contours. Second, the band diagram was computed for an infinite lattice, whereas here, the finite-size platonic lens yields a discrete set of modes, similar to the modes of a rectangular cavity.
Seven resonances associated with seven spots in the k-space are identified (Fig.~\ref{Fig3}). Four of them, $P_{i \in [ 1,2,3,4 ]}$ couple to propagating waves of the plain plate. Three modes with higher transverse k-components, $E_{i \in [ 1,2,3 ]}$ correspond to modes of the lens fed by evanescent waves. As time progresses, energy is released from these modes respectively into propagative and evanescent waves. However, we find that the portion of evanescent components is small relative to the propagative components. This is demonstrated in Fig.~\ref{Fig4} where the lateral field distribution (blue circles) is accurately reproduced by the coherent sum of the fields radiated by modes $P_{i \in [ 1,2,3,4 ]}$ only (red line). Modes $E_{i \in [ 1,2,3 ]}$ do not contribute significantly to the pulse reconstruction which cannot be therefore explained in terms of evanescent wave enhancement \cite{Gomez-Santos,Chan05,Pendry11,Greffet}.

The mechanism involved here is therefore of a different nature. The amplitude and phase of each propagative contribution issued from $P_{i \in [ 1,2,3,4 ]}$ can be represented by a phasor in the complex plane and the total field by a phasor sum. In this representation each contribution rotates at its own speed, as it oscillates freely in time at its own resonant frequency: its amplitude as well as relative phase accumulated during propagation evolve in time. Depending on their relative magnitude and phase, the four propagative components add up coherently at the focal plane to form a spot with size above (Fig.~\ref{Fig4}a), at (Fig.~\ref{Fig4}b), or below (Fig.~\ref{Fig4}c) diffraction limit. At time $t_{\rm focus}$, the resulting field oscillates faster than the sinc function in the vicinity of the focal spot, resulting in sub-wavelength resolution. This super-oscillation is similar to the super-oscillating functions discussed in \cite{Berry&Popescu}.
In our case however, there is no need to design such a function: Time-dependent excitation naturally generates a super-oscillating function when all field components spontaneously synchronize. This spontaneous phase arrangement is not possible with stationary excitation as the oscillation frequency is forced at $f_0$: All phasors are forced to rotate at the same speed with the same amplitude and fixed phase delay. In actuality, the super-oscillatory function can be written explicitly from the knowledge of the transverse k-vector $k_y(P_i)$ and amplitude $A(P_i)$ at the focal point of each mode $P_i$ by replacing each transverse component by a cosine function: $S(y)=\sum_{i=1}^4A(P_i).\cos(k_y(P_i).y)$. Amplitudes $A(P_i)$ are time dependent and are given by the histograms in Fig.~\ref{Fig4}, while the transverse k-vector is measured at the maximum of the mode spot in Fig.~\ref{Fig3}a. This rather simple function reproduces to a good approximation the super-oscillations of the field at $t_{\rm focus}$ as shown in Fig.~\ref{Fig4}c (red circles). The increasing discrepancy away from $y=0$ is attributed to the finite transverse extension of the lens which cannot be rendered by infinitely-extended cosine functions.

Finally, we observe the quantum revival of the super-oscillating function predicted in \cite{Berry&Popescu}. This is seen numerically and shown in the inset of Fig~\ref{Fig2}: Three successive episodes of increase and decrease of the lateral resolution are observed before the signal fades away. The phases self-organize periodically with a period of 600 $\mu$s which corresponds to the beating between modes $P_1$ and $P_3$.

In summary, we have investigated the dynamics of pulse super-focusing by a Veselago-Pendry lens and demonstrated in a simple experiment how the focus resolution improves over time, as predicted theoretically \cite{Chan05,Greffet}. An original mechanism is proposed and experimentally demonstrated, based on genuine spontaneous super-oscillations of the field radiated by a discrete set of well identified lens modes. This explanation is new in the context of flat lens focusing. In contrast to earlier works, the flat lens which we investigated focuses in the acoustic band and does not require a negative effective index \cite{Joannopoulos02}. Another fundamental difference is that evanescent waves do not play any role: Our observations remain valid for thick flat lenses as well as for a source distant from the lens interface. Super-oscillation-based super-focusing depends weakly on the source position, a major advantage over near-field super-focusing. This study raises a number of fundamental issues in the dynamics of waves in metamaterials, e.g. what is the time necessary for the all-angle negative refraction to establish itself, how long does it take for the subwavelength focusing to be reached, various questions that could be easily addressed with our experimental system \cite{Ramakrishna07}. But this study also opens new routes to envision a controlled engineering of flat lens: Tailoring the flat lens resonances and their dynamical interaction should lead to optimized super-focusing.

M.D. acknowledges Ph.D. funding from the Direction G\'{e}n\'{e}rale de l'Armement (DGA).
P.S. is thankful to the Agence Nationale de la Recherche for its support under grant ANR PLATON (No. 12-BS09-003-01), the LABEX WIFI (Laboratory of Excellence within the French Program "Investments for the Future") under reference ANR-10-IDEX-0001-02 PSL* and the Groupement de Recherche 3219 MesoImage.


\end{document}